\documentclass[numberedappendix,onecolumn]{emulateapj}

\usepackage{apjfonts}

\newcommand{\himpc}{{\hbox {$\,h^{-1}$}{\rm Mpc}}}
\newcommand{\higpc}{{\hbox {$\,h^{-1}$}{\rm Gpc}}}
\newcommand{\simgt}{\hspace{0.3em} \raisebox{0.4ex}{$>$}
  \hspace{-0.75em} \raisebox{-.7ex}{$\sim$} \hspace{0.3em}}
\newcommand{\simlt}{\hspace{0.3em} \raisebox{0.4ex}{$<$}
  \hspace{-0.75em} \raisebox{-.7ex}{$\sim$} \hspace{0.3em}}

\begin{document}

\title{An Analysis of the Large Scale $N$-body Simulation using the
Minkowski Functionals}  

\author{Takamichi Nakagami, Takahiko Matsubara}
\affil{Department of Physics and Astrophysics, Nagoya University,
  Chikusa, Nagoya 464-8603} 
\email{nakagami@a.phys.nagoya-u.ac.jp}
\email{taka@a.phys.nagoya-u.ac.jp}

\author{Jens Schmalzing}
\affil{Ludwig-Maximilians-Universit\"at, 
  Theresienstra\ss e 39, 80333 M\"unchen, Germany} 
\email{j.s@lmu.de}

\and

\author{Yipeng Jing}
\affil{Shanghai Astronomical Observatory, the Partner Group of
  Max-Planck-Institut f\"ur Astrophysik, Nandan Road 80, Shanghai
  200030, China}
\email{ypjing@center.shao.ac.cn}


\begin{abstract}
We analyze the Minkowski functionals with a large $N$-body simulation
of a standard $\Lambda$CDM model, focusing on transition scales
between linear and non-linear gravitational evolution. We numerically
calculate the Minkowski functionals with sufficient accuracies to
investigate the transition scales, $10$--$50\himpc$. The results are
compared with analytic formulae of linear and second-order
perturbation theories. We first show that the skewness parameters of
the density fluctuations, which are important in second-order analytic
formulae of the Minkowski functionals, are in good agreement with the
perturbation theory. Considering relative differences between the
Minkowski functionals of the analytic formulae and that of the
simulation data, we evaluate accuracy levels of the predictions of the
perturbation theory. When the straightforward threshold $\nu$ by
density value is used in Minkowski functionals, the accuracy of the
second-order perturbation theory is within $10\%$ for smoothing length
$R > 15\himpc$, and within a several $\%$ for $R > 20\himpc$. The
accuracies of the linear theory are 2--5 times worse than that. When
the rescaled threshold by volume fraction, $\nu_{\rm f}$ is used,
accuracies of both linear and second-order theories are within a few
$\%$ on all scales of $10\himpc < R < 50\himpc$.
\end{abstract}

\section{Introduction}

The morphological patterns of the large-scale structure in the
universe are the consequence of the dynamical evolution of the
universe, and thus provide important information on cosmology. The
pattern of the clustering of galaxies is described in many ways. For
example, \citet{soneira78} indicated the picture of the hierarchical
clustering, and \citet{joeveer86} indicated that of the cell
structure. Gott, Melott, \& Dickinson (1986) (hereafter GMD) proposed
a new picture, which is depicted by sponge-like topology of the
universe. To capture the topology of the universe, it is useful to
look at the three dimensional views of galaxy distributions (Einasto
\& Miller 1983). The clustering pattern is so complicated that it is
difficult to fully describe the pattern by either the hierarchical
picture or the cell picture alone.

GMD showed that the large-scale structure of the universe has
sponge-like topology by analyzing the CfA1 catalog of galaxies. They
examined various cosmological models of the galaxy distribution and
quantified the topology by the genus statistic. The CfA1 catalog which
GMD used is the redshift survey of about 1,800 galaxies, which is
relatively small catalog with respect to the today's standard. After
that survey, numerous redshift surveys such as the Las Campanas
Redshift Survey (LCRS), the IRAS Point Source Catalog Redshift Survey
(PSCz), etc.~have been carried out. Recently, the 2-degree Field
Galaxy Redshift Survey (2dFGRS) was completed. The Sloan Digital Sky
Survey (SDSS) is now in progress. In the SDSS, redshifts of about
$8\times10^5$ galaxies are being observed. These surveys are large
enough to quantitatively discuss the topology of the large-scale
structure on large scales. Topological analysis is one of the major
methods to analyze these redshift surveys (see, e.g., Vogeley et al.
1994; Colley 1997; Canavezes et al. 1998; Colley et al. 2000).

The Minkowski Functionals (MFs) are also used as descriptors of the
morphology of the large-scale structure. While they are originally
mathematical quantities, Mecke, Buchert, \& Wagner (1994) applied the
MFs to the analysis of the galaxy clustering. Schmalzing \& Buchert
(1997) provided a computational algorithm to calculate the MFs. The
genus statistic is one of the MFs in a certain condition.

The MFs possess complementary information to the two-point correlation
function or the power spectrum, which are the most fundamental tools
to quantify the clustering pattern. One of the important applications
of the genus statistic and the MFs is a Gaussianity test of the
primordial density field, which can not be performed by the two-point
correlation function or the power spectrum. Any Gaussian field has a
universal form of the genus statistic and the MFs as functions of the
density threshold. As a result, any deviation from that form indicates
the non-Gaussianity of a density field.

\citet{hikage2003a} analyzed the early SDSS sample using the MFs. They
found the MFs of the SDSS are consistent with that of the $\Lambda$CDM
model. However, their analysis is limited by the cosmic variance
because of the smallness of the data, and the discrimination of models
with various parameters is still difficult. Complete SDSS data would
distinguish different models in more detail.

When we compare observations with theories, the $N$-body simulation
plays an important role. A cosmological $N$-body simulation was first
carried out by Miyoshi \& Kihara (1975). Since then, the number of
particles in $N$-body simulations has been growing every year.
Nowadays simulations with the number of particles of over $10^9$ and
the physical size of $1\ h^{-1}{\rm Gpc}^3$ have come to be
available. Springel et al. (1998) compared the genus calculated from
an $N$-body simulation, which contains $256^3$ particles in a
$(240 h^{-1}{\rm Mpc})^3$ box, with the data of the IRAS 1.2-Jy
redshift survey. \citet{hikage2002} used an $N$-body simulation of the
$\Lambda$CDM model carried out by Jing \& Suto (1998), which contained
$256^3$ particles in a $(300h^{-1}{\rm Mpc})^3$ box. They found that
the genus statistic of their $N$-body simulations is consistent with
the one calculated from the data of the SDSS Early Data Release.
Recently, \citet{hikage2003b} used the Hubble volume simulation with a
box size of $(3000h^{-1}\mbox{Mpc})^3$ and a number of particles $N =
10^9$, and investigate the genus statistics with respect to the
biasing effects.

A drawback of using the $N$-body simulation in comparing theories and
observations is that the $N$-body simulation is computationally costly
to obtain theoretical predictions of given cosmological models.
Fortunately, there are analytic approximations for MFs. The MFs of a
random Gaussian field is well-known (Tomita 1986). However, the
density field in the universe is not exactly random Gaussian even if
the primordial density field is Gaussian, because of the nonlinear
evolution effects. Matsubara (2003) derived the analytic
approximations of the MFs of weakly non-Gaussian fields, and obtained
analytic formulae of the MFs with effects of the weakly nonlinear
evolution by applying the second-order perturbation theory. Since in
the last formulae is assumed an approximation that the non-Gaussianity
is weak, there is a regime that the formulae can be applied.
Identifying such regime is indispensable to use the analytic formulae
in analyses of the observations, instead of performing computationally
costly simulations model by model.

In this paper, we analyze a large $N$-body simulation of the
large-scale structure using the MFs and investigate the transition
scales between linear and non-linear evolution, and compare them with
the theoretical formulae. The accuracy levels of the above analytic
formulae are identified. We use an $N$-body simulation of $512^3$
particles in a box of $(1024h^{-1}{\rm Mpc})^3$.

This paper is organized as follows. In \S 2, we briefly review the
concept of MFs and their theoretical formulae. In \S 3, we explain the
computational methods of the MFs. In \S 4, we analyze the MFs of the
simulation data and compare them with the analytic formulae. In this
section the comparison of the skewness parameters, which play an
important role in weakly nonlinear formulae, is also presented. In \S
5, we summarize the results and discuss future outlook.

\section{The Minkowski Functionals}

\subsection{Definitions}

Mecke, Buchert, \& Wagner (1994) and Schmalzing, \& Buchert (1997)
proposed the MFs as the geometrical descriptors of the distribution of
galaxies. These quantities are originally derived by Minkowski (1903).
To define the MFs of galaxy distributions, we first consider
isodensity contours of the density contrast
$\delta=(\rho-\bar{\rho})/\bar{\rho}$, where $\rho$ is the density
field and $\bar{\rho}$ is the mean density. Then we identify the set
of regions $M$ where the density contrast $\delta$ exceeds some
threshold. We use the threshold $\nu$ which is defined by
$\nu\equiv\delta_{\rm th}/\sigma_0$ for the moment, where $\delta_{\rm
th}$ is the threshold density contrast and
$\sigma_0\equiv\sqrt{\langle\delta^2\rangle}$ is the {\em rms} of the
density contrast. There are $d+1$ MFs in a $d$-dimensional space. The
case $d=3$ is of our primary interest. In this case, the MFs
correspond to the following quantities: (1) the fractional volume
enclosed by contours
\begin{eqnarray}
V_0=\frac{1}{V}\int_MdV,
\end{eqnarray}
where $V$ is the total volume, (2) the surface area per volume
\begin{eqnarray}
V_1=\frac{1}{6V}\int_{\partial M}d^2A,
\end{eqnarray}
where the region of the integral $\partial M$ is the surfaces of the
contours, (3) the integrated mean curvature
\begin{eqnarray}
V_2=\frac{1}{6\pi V}\int_{\partial
  M}\left(\frac{1}{R_1}+\frac{1}{R_2}\right)d^2A,
\end{eqnarray}
where $1/R_1$ and $1/R_2$ are the principal curvatures on the surface,
and (4) the integral Gaussian curvature
\begin{eqnarray}
V_3=\frac{1}{4\pi V}\int_{\partial
  M}\frac{1}{R_1R_2}d^2A.
\end{eqnarray}
The last quantity is proportional to the Euler characteristic of the
contour surfaces. The Euler characteristic is also proportional to the
genus statistic if the boundary of the sample is neglected. Further
geometrical meanings of MFs are found in \citet{mecke94}.

\subsection{Prediction for First- and Second-order Perturbations}

\subsubsection{Gaussian random field}

The analytic formulae of the MFs as functions of the threshold $\nu$
for random Gaussian field is given by (Tomita 1986; Schmalzing \&
Buchert 1997):
\begin{eqnarray}
  V_k(\nu) = 
  \frac{1}{(2\pi)^{(k+1)/2}}
  \frac{\omega_3}{\omega_{3-k}\omega_k}
  \left(\frac{\sigma_1^2}{3\sigma_0^2}\right)^{k/2}
  H_{k-1}(\nu)e^{-\nu^2/2},
\label{eq:48}
\end{eqnarray}
where $k=0,1,2,3$, and the factor $\omega_k$ is the volume of the unit
ball in $k$-dimensions, i.e., $\omega_0=1$, $\omega_1=2$,
$\omega_2=\pi$, and $\omega_3=4\pi/3$.
The quantities $\sigma_0^2$ and $\sigma_1^2$ are defined by
\begin{eqnarray}
  \sigma_0^2\equiv\langle\delta^2\rangle,\qquad
  \sigma_1^2\equiv\langle(\nabla\delta)^2\rangle,
\label{eq:46}
\end{eqnarray}
and the functions $H_n (n=0,2,3,\cdots)$ are the Hermite polynomials
with a convention,
\begin{eqnarray}
  H_n(\nu) =
  e^{\nu^2/2}\left(-\frac{\partial}{\partial\nu}\right)^ne^{-\nu^2/2}.
\label{eq:49}
\end{eqnarray}
Following Matsubara (2003), we use a notation
\begin{eqnarray}
  H_{-1}(\nu)\equiv
  e^{\nu^2/2}\int^\infty_\nu d\nu e^{-\nu^2/2} =
  \sqrt[]{\mathstrut \frac{\pi}{2}}
   e^{\nu^2/2}{\rm erfc}
  \left(\frac{\nu}{\sqrt[]{\mathstrut 2}}\right),
\label{eq:4a}
\end{eqnarray}
when $k=0$.

\subsubsection{Non-Gaussian field}

In non-Gaussian random fields, analytic formula with effects of the
weakly non-linear evolution are also derived (Matsubara 2003). Using
the Edgeworth-like expansion, the formulae is given by
\begin{eqnarray}
&&
  V_k(\nu) =
  \frac{1}{(2\pi)^{(k+1)/2}}
  \frac{\omega_3}{\omega_{3-k}\omega_k}
  \left(\frac{\sigma_1}{\sqrt[]{\mathstrut 3}\sigma_0}\right)^k
  e^{-\nu^2/2}
%\nonumber \\
%&& \qquad \times
  \left\{
    H_{k-1}(\nu) +
    \left[
      \frac{1}{6}S^{(0)}H_{k+2}(\nu) +
      \frac{k}{3}S^{(1)}H_k(\nu)
%%     \right.
%%   \right.
%% \nonumber \\
%% &&\qquad\qquad
%%   \left.
%%     \left.
    + \frac{k(k-1)}{6}S^{(2)}H_{k-2}(\nu)\right]\sigma_0 +
    {\cal O}(\sigma_0^2)\right\},
\nonumber\\
\label{eq:422}
\end{eqnarray}
where $S^{(a)}$ $(a=0,1,2)$ are the skewness parameters which are
defined by
\begin{eqnarray}
  S^{(0)}&\equiv&\frac{\langle\delta^3\rangle}{\sigma_0^4},
\label{eq:436} \\
  S^{(1)}&\equiv&-\frac{3}{4}
  \frac{\langle\delta^2(\nabla^2\delta)\rangle}{\sigma_1^2\sigma_0^2},
\label{eq:437} \\
  S^{(2)}&\equiv&
  -\frac{9}{4}
  \frac{\langle(\mbox{\boldmath $\nabla$} \delta \cdot 
    \mbox{\boldmath $\nabla$}\delta)\nabla^2\delta\rangle}{\sigma_1^4}.
\label{eq:438}
\end{eqnarray}
The above formula is a general one for weakly non-Gaussian fields.

In second-order perturbation theory, the skewness parameters of a
smoothed density field, convolved with a Gaussian smoothing kernel are
given by
\begin{eqnarray}
S^{(0)}(R)&=&(2+E)S^{11}_0-3S^{02}_1+(1-E)S^{11}_2,
\label{eq:439} \\
S^{(1)}(R)&=&\frac{3}{2}\left[\frac{5+2E}{3}S^{13}_0-\frac{9+E}{5}S^{22}_1-S^{04}_1+\frac{2(2-E)}{3}S^{13}_2-\frac{1-E}{5}S^{22}_3\right],
\label{eq:440} \\
S^{(2)}(R)&=&9\left[\frac{3+2E}{15}S^{33}_0-\frac{1}{5}S^{24}_1-\frac{3+4E}{21}S^{33}_2+\frac{1}{5}S^{24}_3-\frac{2(1-E)}{35}S^{33}_4\right],
\label{eq:441}
\end{eqnarray}
where the factor $S^{\alpha\beta}_m(R)$ is defined by
\begin{eqnarray}
  S^{\alpha\beta}_m(R) &\equiv&
  \frac{\sqrt[]{\mathstrut 2\pi}}{\sigma_0^4}
  \left(\frac{\sigma_0}{\sigma_1R}\right)^{\alpha+\beta-2}
%\nonumber \\
%&&\times
  \int\frac{l_1^2dl_1}{2\pi^2R^3}
  \frac{l_2^2dl_2}{2\pi^2R^3} P_{\rm lin}
  \left(\frac{l_1}{R}\right)P_{\rm lin}
  \left(\frac{l_2}{R}\right)
  e^{-l_1^2-l_2^2}l_1^{\alpha-3/2}l_2^{\beta-3/2}I_{m+1/2}(l_1l_2),
\label{eq:442}
\end{eqnarray}
and $R$ is a smoothing length, $I_\nu(z)$ is the modified Bessel
function. In linear theory, the variances $\sigma_0^2$ and
$\sigma_1^2$ are given by
\begin{eqnarray}
\sigma_j^2(R)=\int\frac{k^2dk}{2\pi^2}k^{2j}P_{\rm lin}(k)W^2(kR),
\label{eq:443}
\end{eqnarray}
with $j=0,1$. In this paper we only consider the Gaussian smoothing
kernel $W_R(x)=\pi^{-3/2}R^{-3} \exp(-x^2/2R^2)$ and thus the window
function is given by $W(kR) = e^{-(kR)^2}$ which is a Fourier
transform of the smoothing kernel. The notation $P_{\rm lin}(k)$
indicates the linear power spectrum. The constant $E$ depends on the
cosmological parameters and is approximately equal to 3/7 (Matsubara
2003) for many sensible models. The values of the skewness parameters
are independent on the amplitude of the linear power spectrum in a
lowest-order approximation. The linear power spectrum is given by the
adiabatic CDM model with Harrison-Zel'dovich spectrum (Bardeen et al.
1986). Neglecting other components, such as baryons and neutrinos, the
skewness parameters $S^{(a)}(a=0,1,2)$ are functions of only a
combination $\Gamma R$ where $\Gamma$ is the shape parameter of the
CDM transfer function. The shape parameter $\Gamma=0.2$ is adopted in
our $N$-body simulation of this paper. In Table \ref{tab:63}, the
values of $\sigma_0$, $\sigma_1$ in linear theory, and the skewness
parameters in second-order perturbation theory are shown for various
smoothing lengths, assuming the same cosmological parameters as that
of the simulation.
\begin{table}
\begin{center}
\caption{The analytical values of $\sigma_0$, $R\sigma_1/\sigma_0$ and
the skewness parameters $S^{(a)}$ calculated in the CDM model. The
shape parameter $\Gamma=0.2$ and an approximation $E=3/7$ are adopted.
The power spectrum is normalized by $\sigma_8=0.9$.}
\label{tab:63}
\begin{tabular}{crrrrrrrrr}  \hline\hline
$R (h^{-1}{\rm Mpc})$ &10 & 12 & 15 & 18 & 20 & 25 & 30 & 40 & 50 \\
$\Gamma R$& 2.0 & 2.4 & 3.0 & 3.6 & 4.0 & 5.0 & 6.0 & 8.0 & 10.0 \\ \hline
$\sigma_0$ & 0.394 & 0.328 & 0.258 & 0.210 & 0.185 & 0.141 & 0.111 &
0.075 & 0.054 \\
$R\sigma_1/\sigma_0$ & 0.990 & 1.018 & 1.052 & 1.079 & 1.095 & 1.127 &
1.153 & 1.191 & 1.220 \\
$S^{(0)}$ & 3.500 & 3.453 & 3.398 & 3.355 & 3.332 & 3.285 & 3.250 &
3.201 & 3.169 \\
$S^{(1)}$ & 3.566 & 3.514 & 3.453 & 3.404 & 3.377 & 3.324 & 3.284 &
3.228 & 3.191 \\
$S^{(2)}$ & 3.662 & 3.668 & 3.679 & 3.692 & 3.701 & 3.723 & 3.744 &
3.783 & 3.818 \\ \hline
\end{tabular}
\end{center}
\end{table}

\begin{figure}
\epsscale{0.9}
\plotone{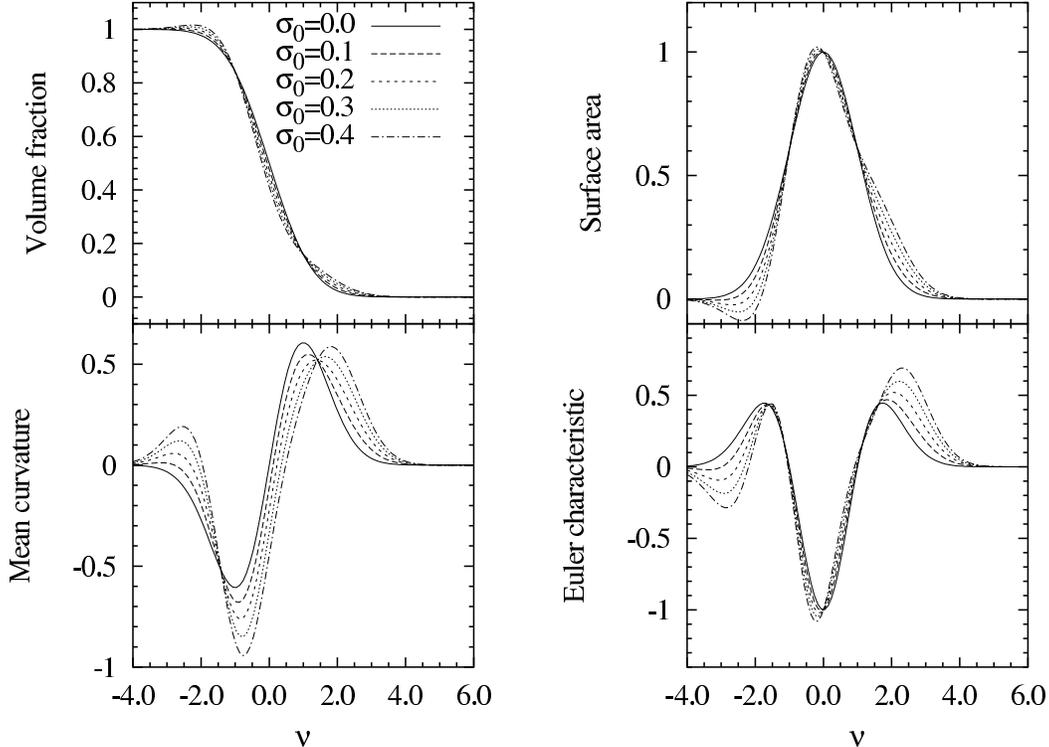}
\caption{The MFs for Gaussian and weakly non-Gaussian random fields.
The solid lines with label $\sigma_0 = 0$ correspond to a Gaussian field.
Other curves with $\sigma_0=0.1, 0.2, 0.3, 0.4$ correspond to the weakly
non-Gaussian field.}
\label{fig:41}
\end{figure}

In Figure \ref{fig:41}, we plot the functions of equation
(\ref{eq:422}).
The skewness parameters in this Figure are calculated
by assuming $\Gamma R=4.0$. The curves of $\sigma_0=0.1, 0.2, 0.3,
0.4$ are depicted. The curves with the label $\sigma_0=0.0$ correspond
to the Gaussian limit $\sigma_0 \rightarrow 0$.

\subsection{Conversion of the Threshold Density by Volume Fractions}

Although we have used the threshold density $\nu=\delta_{\rm
th}/\sigma_0$ so far, we can use another definition of the threshold
$\nu_{\rm f}$ (GMD) by the volume fraction of the high-density
regions:
\begin{eqnarray}
  V_0 =
  \frac{1}{\sqrt[]{\mathstrut 2\pi}}
    \int^\infty_{\nu_{\rm f}}dx e^{-x^2/2}.
\label{eq:433}
\end{eqnarray}
In random Gaussian fields, the two definitions of $\nu$ and $\nu_{\rm
f}$ are identical. In most of the previous work on the genus analysis,
the rescaled threshold $\nu_{\rm f}$ is commonly employed. This
re-scaling removes exactly the effect of the evolution of the
one-point probability distribution of the density field. Since the
two-point characteristics of the density field only affect the MF
curves through the parameters $\sigma_0$ and $\sigma_1$, the remaining
differences must be due to higher-order characteristics.

Analytic formulae of the MFs by the threshold $\nu_{\rm f}$ is given
by (Matsubara 2003)
\begin{eqnarray}
  V_k(\nu_{\rm f}) &=&
  \frac{1}{(2\pi)^{(k+1)/2}}
  \frac{\omega_3}{\omega_{3-k}\omega_k}
  \left(\frac{\sigma_1}{\sqrt[]{\mathstrut 3}\sigma_0}\right)^k
%\nonumber \\
%&& \times
  e^{-\nu_{\rm f}^2/2}
  \left\{H_{k-1}(\nu_{\rm f}) +
  \left[\frac{k}{3}(S^{(1)}-S^{(0)})H_k(\nu_{\rm f})
%  \right.\right.
%\nonumber \\
%&& \hspace{3cm}\left.\left.
  +
  \frac{k(k-1)}{6}(S^{(2)}-S^{(0)})H_{k-2}(\nu_{\rm f})
  \right]\sigma_0
  +{\cal O}(\sigma_0^2)\right\}.
\label{eq:435}
\end{eqnarray}
The highest-order Hermite polynomial in the weakly non-Gaussian terms
vanishes in each case. Moreover, the skewness parameters only appear
in the form of differences, $S^{(a)}-S^{(0)}$ $(a=1,2)$. Therefore,
the resultant equation with $\nu_{\rm f}$ is simpler than the original
form using the threshold $\nu$. Because the values of
skewness parameters are usually close, the non-Gaussian correction is
small with the rescaled threshold $\nu_{\rm f}$.

\section{Computational Methods}

Our $N$-body simulation is similar to the one carried out by Jing, \&
Suto (2002), but the physical box size $1024h^{-1}{\rm Mpc}$ is much
larger than that. The number of particles is the same, $N=512^2$. This
simulation was carried out with the particle-particle-particle-mesh
(P$^3$M) code on the vector-parallel machine VPP5000 at the National
Astronomical Observatory of Japan.

The assumed cosmological model is a flat CDM model with a
cosmological constant ($\Omega_0=0.3$, $\lambda_0=0.7$). The
primordial density fluctuations are assumed to be random Gaussian, and
the power spectrum is given by the Harrison-Zel'dovich spectrum. The
linear transfer function for the power spectrum of the dark matter is
that for adiabatic CDM fluctuations. We use the fitting formula by
\citet{bardeen86} with the shape parameter $\Gamma=\Omega_0h=0.2$. The
amplitude of the power spectrum is determined by $\sigma_8=0.9$ at the
present time. Table \ref{tab:62} shows the parameters in this
simulation.
\begin{table}
\begin{center}
\caption{Model parameters of our $N$-body simulation.}
\label{tab:62}
\begin{tabular}{cccccccc}  \hline\hline
Model & No. of particles & Physical box size & $\Omega_0$ &
$\lambda_0$ & $h$ & $\sigma_8$ & $\Gamma$ \\ \hline
$\Lambda$CDM & $512^3$ & $1024h^{-1}{\rm Mpc}$ & 0.30 & 0.70 & 0.6667
& 0.90 & $0.2$ \\ \hline
\end{tabular}
\end{center}
\end{table}

In studying MFs, we first generate a continuous density field of
galaxies. In order to handle the continuous field, we set the grids in
the simulation box. Then the field smoothing is applied in Fourier
space, since the direct convolution in real space is much slower to
compute. The larger the grid number is, the more precisely the density
field is approximated. However, the large number of points requires a
long time to compute. We need to find an appropriate number of grid
points, keeping the calculation as accurate as possible.

\begin{figure}
\epsscale{0.9}
\begin{center}
\plotone{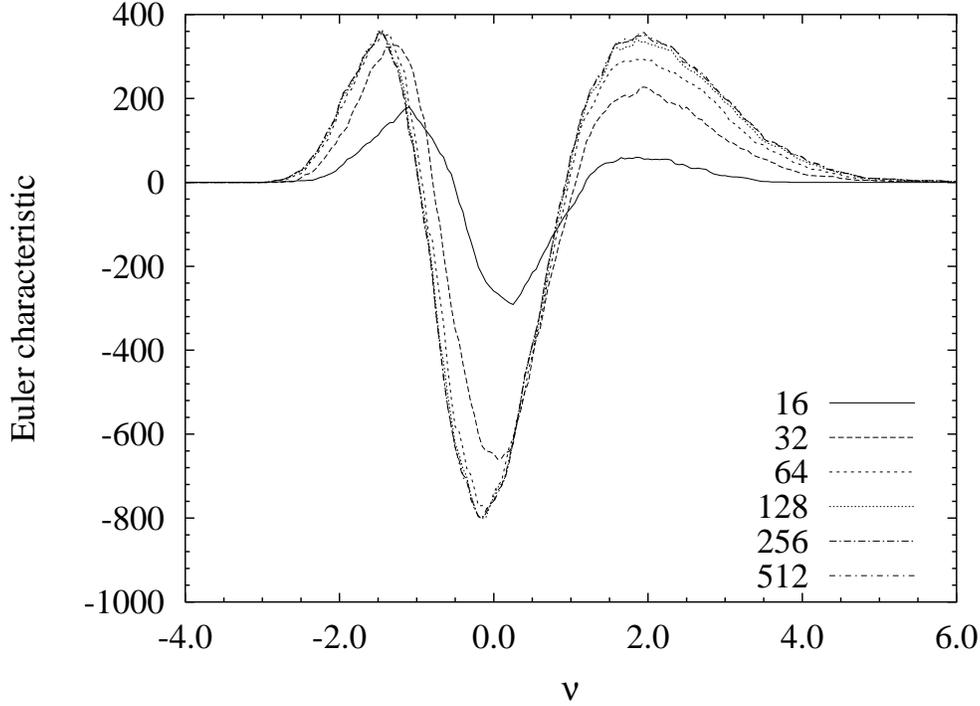}
\end{center}
\caption{The dependence of Euler characteristic on the number of grid
points. The smoothing length is fixed to $20h^{-1}{\rm Mpc}$.
One-dimensional grid numbers are 16, 32, 64, 128, 256 and 512 as
indicated in the figure. The curve does not converge and behave
awkwardly when we adopt the grid number of 16 or 32. On the other
hand, the curves of 128, 256, 512 are difficult to distinguish in the
figure, thus the result is converged.}
\label{fig:51}
\end{figure}

In Figure \ref{fig:51}, the Euler characteristic calculated from the
$N$-body simulation are plotted.
The smoothing length is fixed to
$20h^{-1}{\rm Mpc}$, and the number of grid is varied from $16^3$ to
$512^3$. The values of the Euler characteristic converge when the grid
number is large as expected. Since $256^3$ and $512^3$ grids
essentially give the same result, we consider results with $256^3$
grids are accurate enough. The $128^3$ grids already give reasonable
approximation. Similar tendencies are seen in other MFs. We conclude
that the grid number is sufficient when the spacing of the grids is
larger than $R/2.5$. Therefore, as long as $R \simgt 10\himpc$,
$256^3$ grids in our simulation box $(1024\himpc)^3$ are enough.

Next, we explain error estimation methods in calculating the MFs from
the data of the $N$-body simulation. The comparison between the
analytic formulae in weakly non-linear regions and the MFs calculated
from the $N$-body simulation can be properly done only if we estimate
the correct errors in the simulation. There is a constraint in this
work that we have only one set of the $N$-body simulation data. If
there is only one set of data, we can obtain only one set of the MFs.
To calculate the errors, first we divide the cubic box of the $N$-body
simulation into eight sub-cubes, and then obtain the values of MFs in
each cubes. The errors are estimated by their averages and variances
divided by a scaling factor, $\sqrt{7}$.

\section{Results}

\subsection{The values of the skewness parameters}

\begin{figure}
\epsscale{0.9}
\begin{center}
\plotone{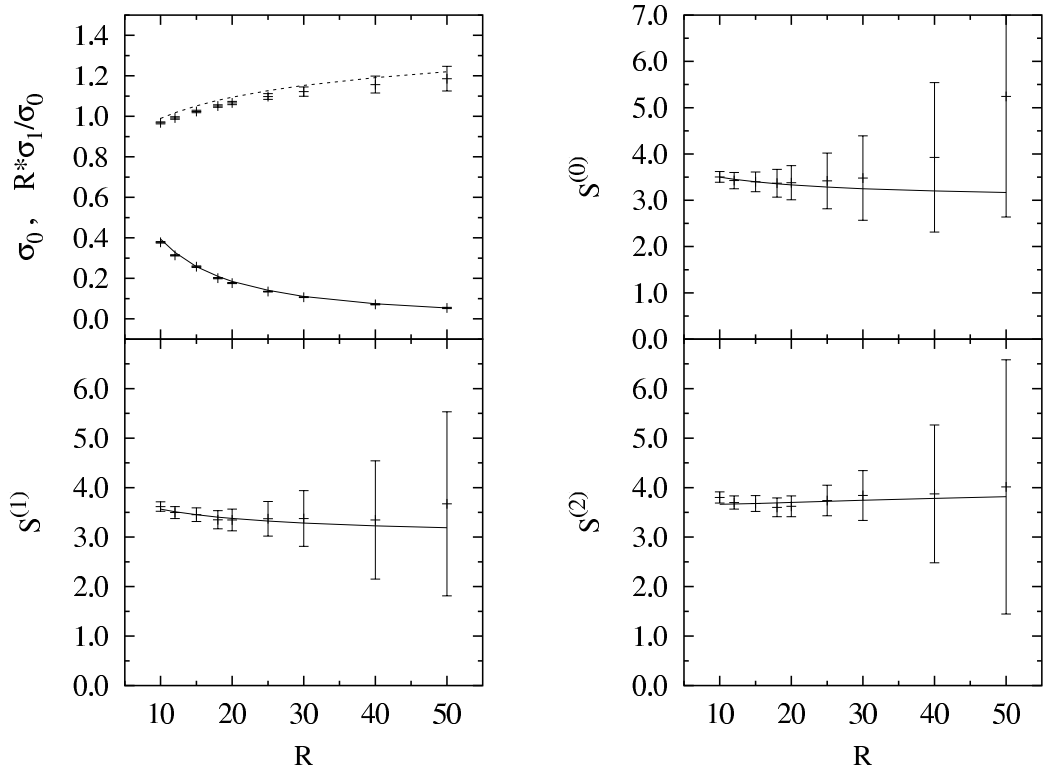}
\end{center}
\caption{The values of $\sigma_0$, $R\times\sigma_1/\sigma_0$ and the
skewness parameters $S^{(a)}$. The analytical values are shown with
lines, and the values of the $N$-body simulation are shown with
symbol. 1 $\sigma$ error bars are also shown for the parameters of the
$N$-body simulation.}
\label{fig:1}
\end{figure}

Before examining the MFs, we first calculate the skewness parameters
from the simulation, since these parameters are essential in analytic
formulae of the MFs. Table \ref{tab:61} and Figure \ref{fig:1} show
the standard deviation $\sigma_0$ of the density distribution of the
galaxy $\delta$, $R\sigma_1/\sigma_0$, and the skewness
parameters of equations (\ref{eq:436})-(\ref{eq:438}) calculated
numerically from the $N$-body simulation.
\begin{table}
\begin{center}
\caption{The values of $\sigma_0$, $R\sigma_1/\sigma_0$ and the
skewness parameters $S^{(a)}$ from numerical calculations of equations
(\ref{eq:436})-(\ref{eq:438}) by the $N$-body simulation.}
\label{tab:61}
\begin{tabular}{cccccc}  \hline\hline
$R (h^{-1}{\rm Mpc})$ & 10 & 12 & 15 & 18 & \\ \hline
$\sigma_0$ &
$0.378\pm0.004$ & $0.314\pm0.003$ & $0.258\pm0.004$ & $0.201\pm0.003$ & \\
$R\sigma_1/\sigma_0$ &
$0.968\pm0.005$ & $0.992\pm0.006$ & $1.024\pm0.006$ & $1.051\pm0.006$ & \\
$S^{(0)}$ &
$3.505\pm0.117$ & $3.425\pm0.175$ & $3.398\pm0.212$ & $3.368\pm0.299$ & \\
\(S^{(1)}\) &
$3.617\pm0.095$ & $3.496\pm0.121$ & $3.453\pm0.138$ & $3.350\pm0.183$ & \\
\(S^{(2)}\) &
$3.800\pm0.113$ & $3.699\pm0.133$ & $3.679\pm0.159$ & $3.601\pm0.190$& \\
\hline
&&&&& \\
\hline\hline
$R (h^{-1}{\rm Mpc})$ & 20 & 25 & 30 & 40 & 50 \\ \hline
\(\sigma_0\) &
$0.177\pm0.003$ & $0.135\pm0.002$ & $0.107\pm0.002$ & $0.072\pm0.003$
& $0.054\pm0.003$ \\
$R\sigma_1/\sigma_0$ & $1.066\pm0.008$ & $1.098\pm0.014$ &
$1.122\pm0.023$ & $1.157\pm0.042$ & $1.186\pm0.061$ \\ 
\(S^{(0)}\) & $3.379\pm0.369$ & $3.418\pm0.603$ & $3.480\pm0.912$ &
$3.926\pm1.641$ & $5.244\pm2.607$ \\ 
\(S^{(1)}\) & $3.347\pm0.220$ & $3.370\pm0.350$ & $3.376\pm0.562$ &
$3.346\pm1.194$ & $3.673\pm1.859$ \\ 
\(S^{(2)}\) & $3.623\pm0.211$ & $3.739\pm0.309$ & $3.841\pm0.504$ &
$3.873\pm1.392$ & $4.015\pm2.569$ \\ \hline 
\end{tabular}
\end{center}
\end{table}
We use $256^3$ grids
commonly for each smoothing length. In Figure \ref{fig:1}, the
analytic predictions and numerical values from the simulation of these
parameters are compared.

As obviously seen in Figure~\ref{fig:1}, most of the figures agree
within the range of numerical errors. There have been similar studies
comparing analytic and numerical values of $S^{(a)}$ and disagreements
are reported \citep[e.g.,][]{colley2000,hikage2003b}. One of the
reason of the disagreements is that those samples are not large
enough in volume and hence the calculations of the skewness parameters
are affected by the cosmic variance. Other reason is that the previous
work uses biased sample in the simulation. Our comparison is directly
made by the dark matter distribution in order to separate the complex
biasing effects and to concentrate on purely nonlinear evolution.

There are some disagreements in the values of $\sigma_0$ and
$\sigma_1$, which are expected if we take into account the fact that
non-linear corrections of order $\sigma_0^2$ should be added to the
linear predictions of $\sigma_0$. Without the second-order correction,
the theoretical values of $\sigma_0$ have a tendency to the
underestimation. This can be understood by the fact that the
correlation function integrated over whole space have to be zero while
the powers on small scales are enhanced by nonlinear evolution.
Consequently, the powers on weakly nonlinear scales are suppressed and
the linear predictions of $\sigma_0$ overestimates the power (Peacock,
\& Dodds 1994, 1996).

\subsection{The comparison of the Minkowski Functionals between the
$N$-body simulation and the analytic formulae}

\begin{figure}
\epsscale{0.9}
\begin{center}
\plotone{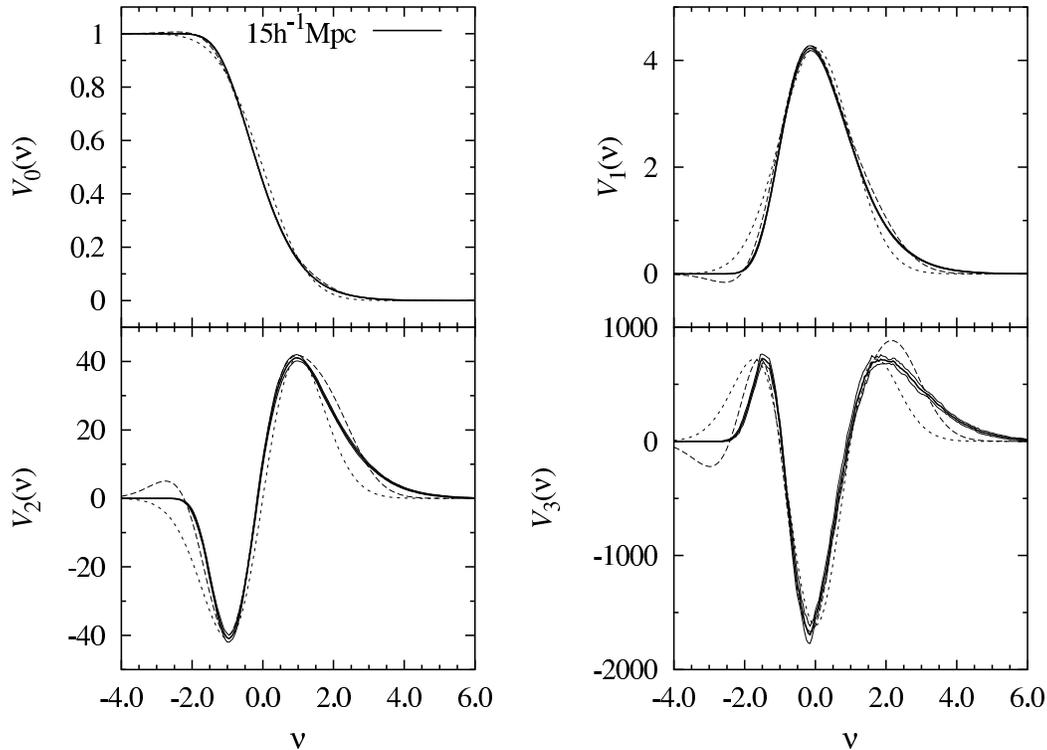}
\caption{The comparison of the MFs for the threshold $\nu$ at the
smoothing length $R=15h^{-1}{\rm Mpc}$. {\it Thick solid lines}: the
MFs calculated from $N$-body simulation; {\it thin solid lines}: 1
$\sigma$ errors; {\it short-dashed lines}: analytical curves for
Gaussian random field; {\it long-dashed lines}: analytical curves for
weakly non-linear evolution. }
\label{fig:6100}
\end{center}
\end{figure}

\begin{figure}
\epsscale{0.9}
\begin{center}
\plotone{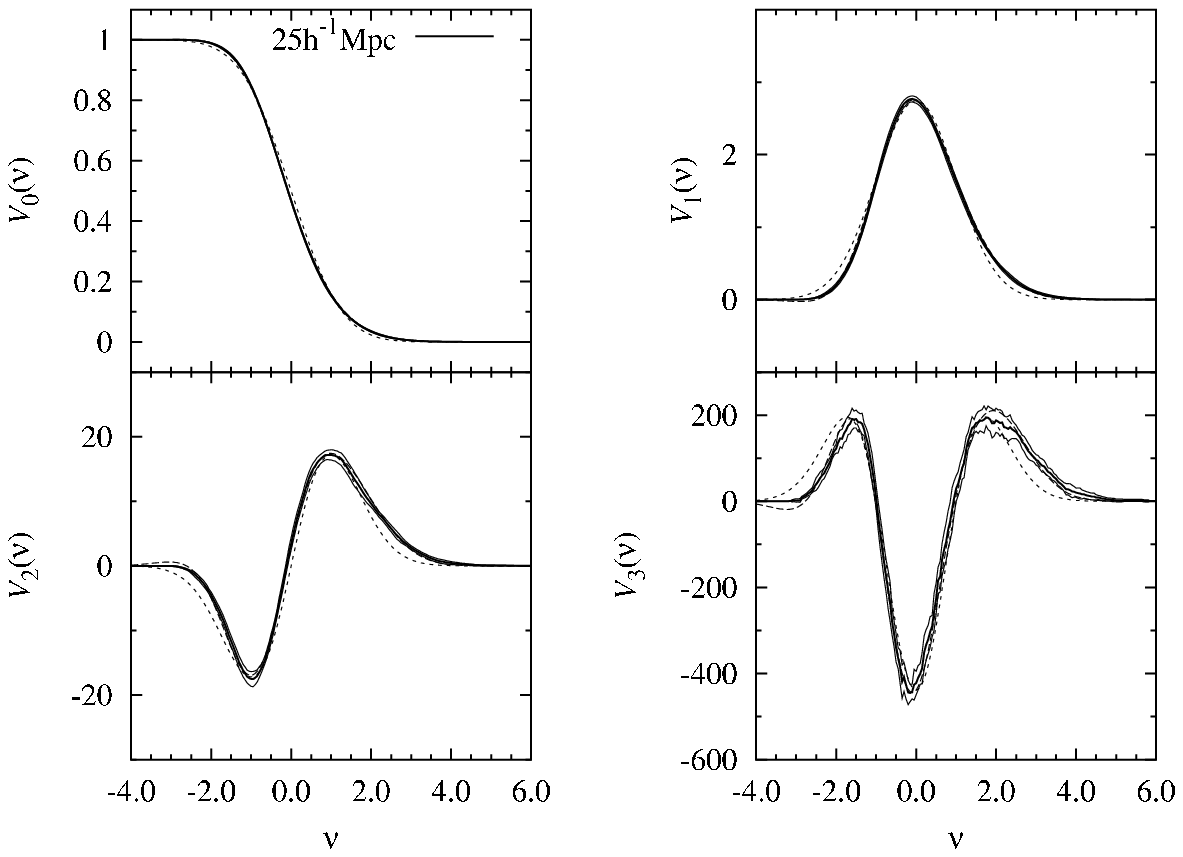}
\caption{Same as figure \ref{fig:6100}, but for the smoothing length 
$R=25h^{-1}{\rm Mpc}$.}
\label{fig:6102}
\end{center}
\end{figure}

\begin{figure}
\epsscale{0.9}
\begin{center}
\plotone{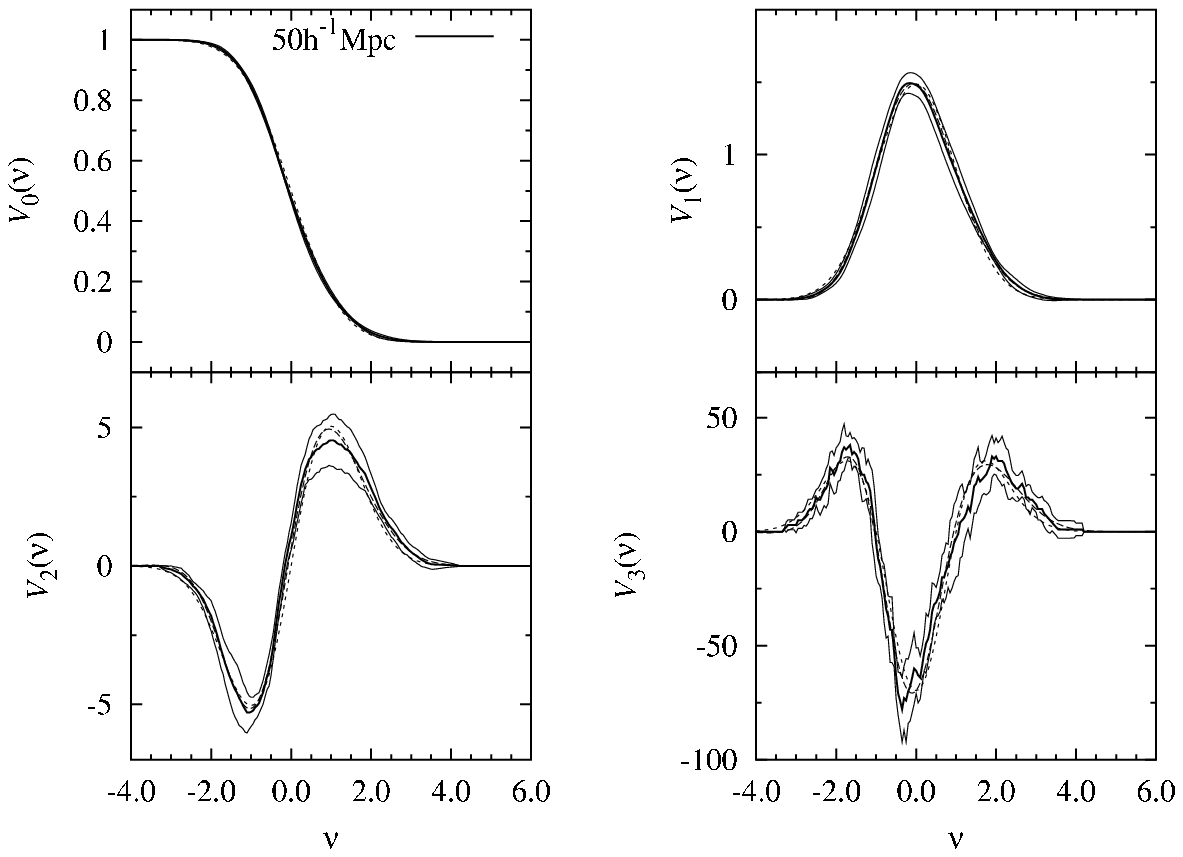}
\caption{Same as figure \ref{fig:6100}, but for the smoothing length 
$R=50h^{-1}{\rm Mpc}$.}
\label{fig:6103}
\end{center}
\end{figure}

In this section, we compare the MFs calculated from the $N$-body
simulation with the analytic formulae. The calculation method of the
MFs of the $N$-body simulation is based on Crofton's formula (Crofton
1868), derived from integral geometry \citep[for details,
see][]{sch97}. Figures \ref{fig:6100}, \ref{fig:6102} and
\ref{fig:6103} show the four MFs $V_k(\nu)$ ($k=0,1,2,3$).
Each MF is
calculated adopting smoothing lengths of 15, 25, and $50h^{-1}{\rm
Mpc}$. The analytic formulae of the Gaussian random fields are plotted
with short-dashed lines, and those of the weakly non-linear fields
with long-dashed lines. The amplitudes of the analytic curves are
estimated by calculating $\sigma_0$ and $\sigma_1$ directly from the
$N$-body simulation.

\begin{figure}
\epsscale{0.9}
\begin{center}
\plottwo{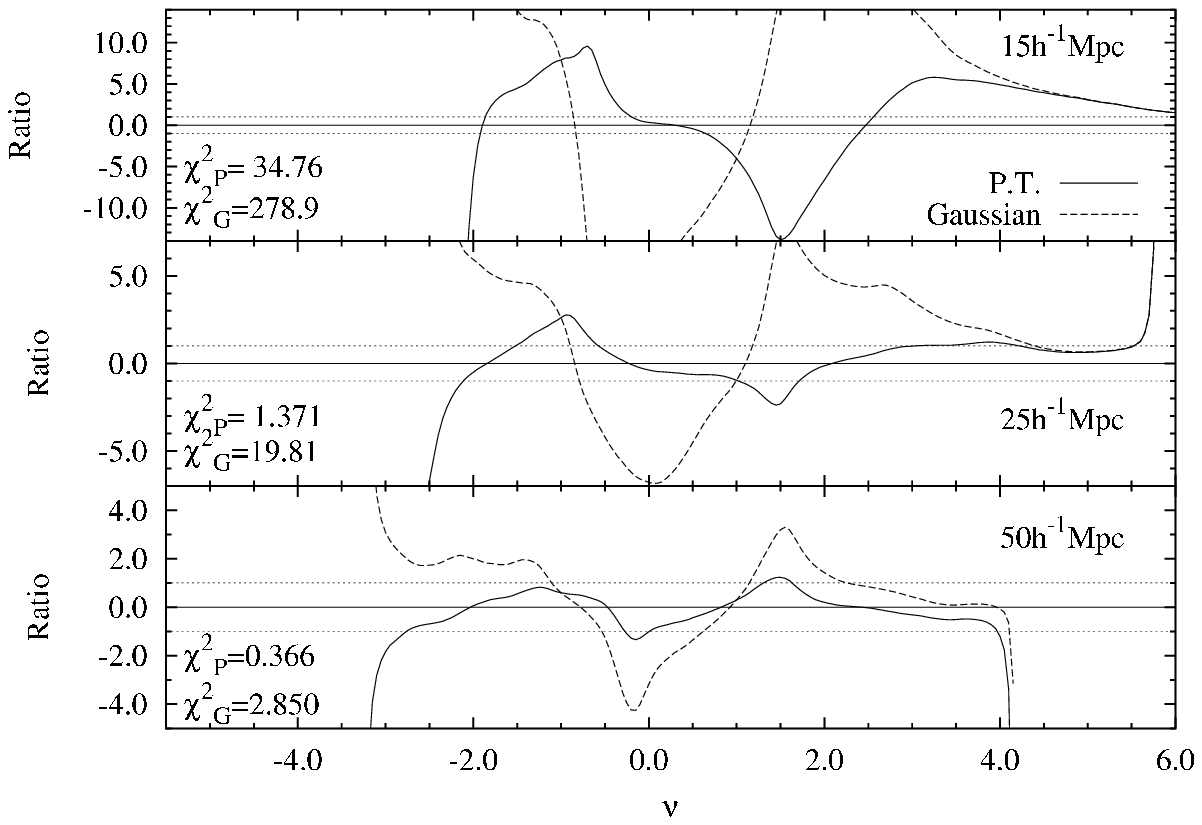}{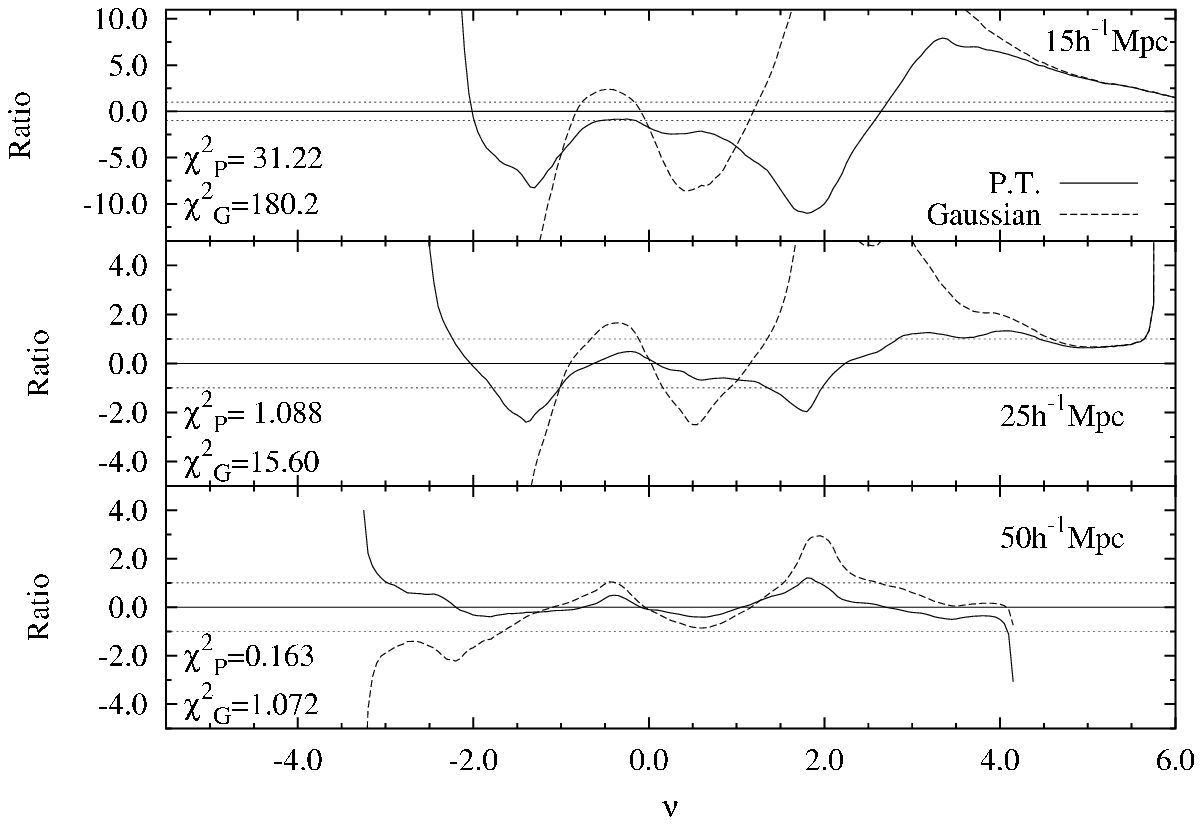}
\plottwo{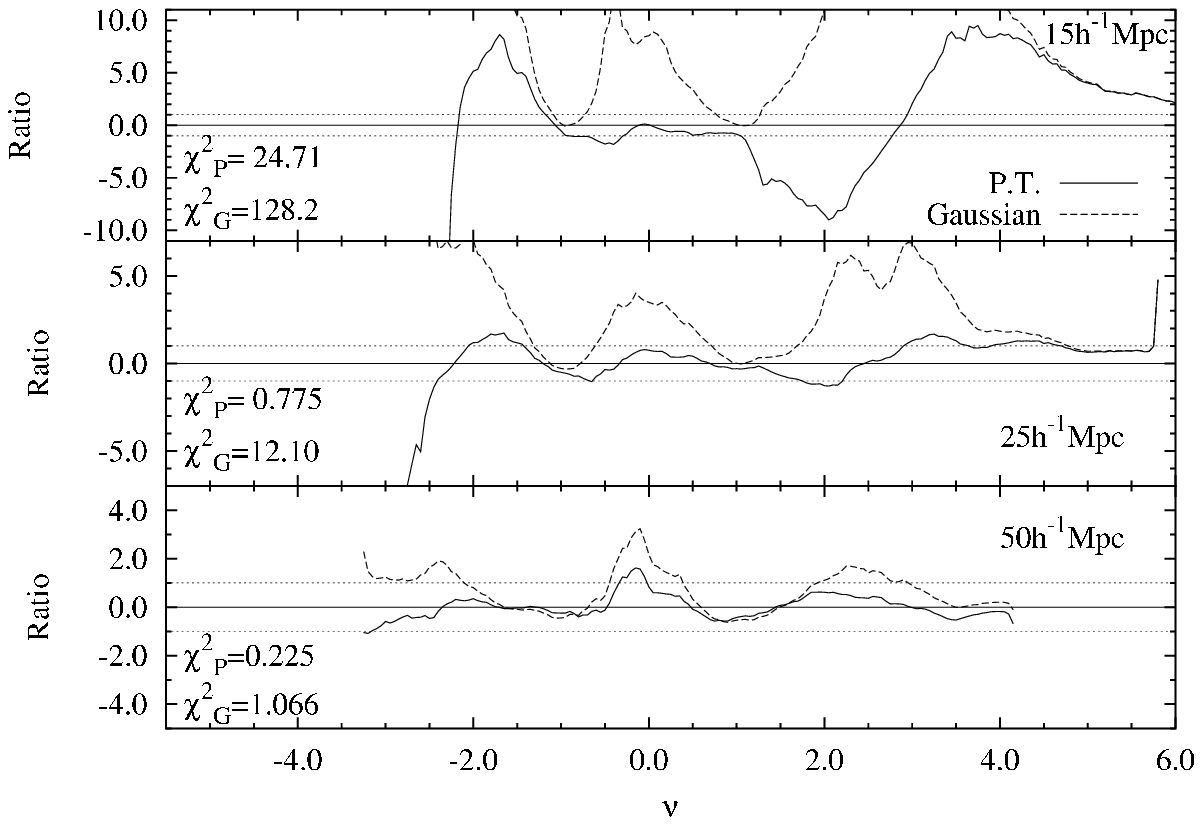}{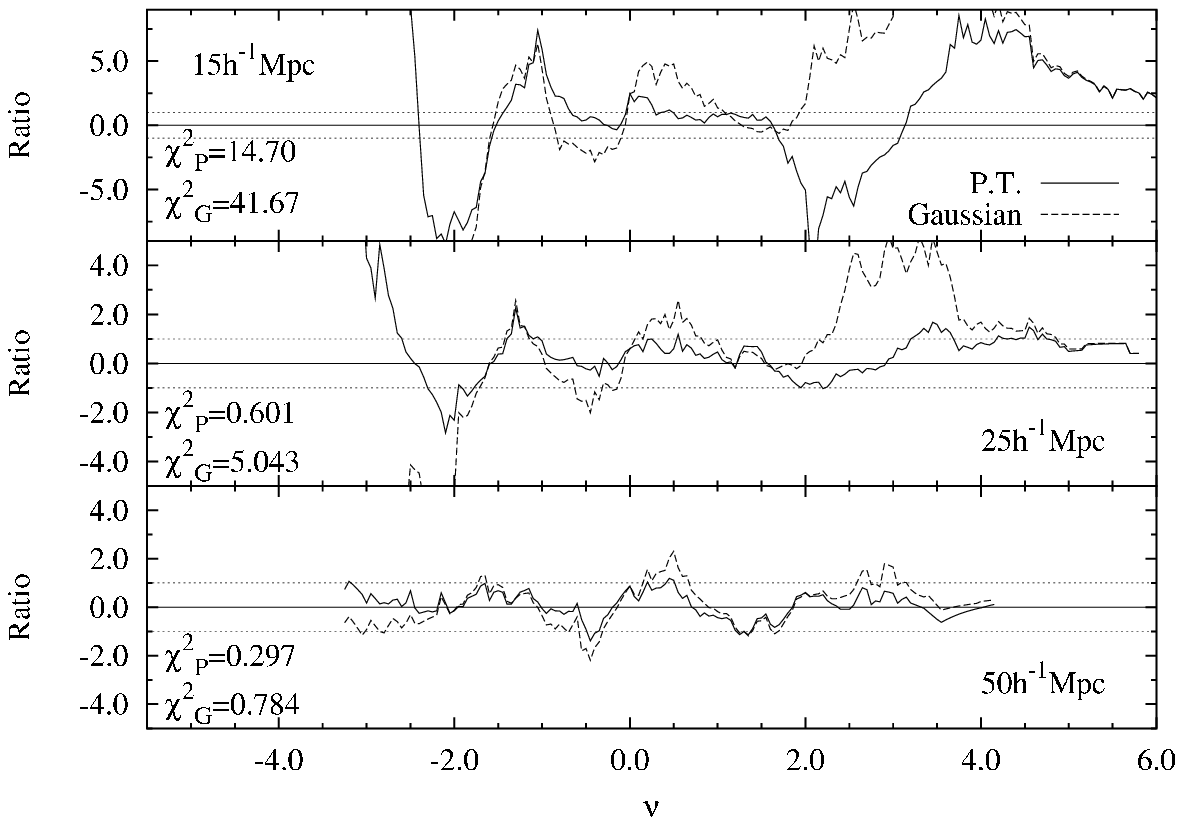}
\caption{The ratio of difference between the MFs of $N$-body simulation
and the analytical values to the error at the smoothing lengths
$R=$15, 25 and 50$h^{-1}{\rm Mpc}$. {\it Upper left panel}: the volume
fraction; {\it upper right panel}: the surface area; {\it lower left
panel}: the integrated mean curvature; {\it lower right panel}: the
Euler characteristic. $\chi^2_{\rm P}/{\rm d.o.f.}$ and $\chi^2_{\rm
G}/{\rm d.o.f.}$ calculated in the range $-2\leq\nu\leq+4$ are also
shown. }
\label{fig:61}
\end{center}
\end{figure}

Figure \ref{fig:61} shows the differences between the MFs of the
$N$-body simulation and that of the analytic formulae.
In the plots,
the differences divided by the $1\sigma$ errors of $N$-body data are
shown. Thus if the curves are in the range of $(-1,+1)$, the two
curves are consistent with 1$\sigma$ significance. As expected, the
MFs calculated from the $N$-body simulation agree with the analytic
formulae for large smoothing lengths.

For quantitative comparisons, we calculate the mean differences
between normalized MF values of the analytic prediction and that of
the simulation:
\begin{eqnarray}
  D_k \equiv
  \frac{1}{\left|V_k(\rm{max})\right|}
  \sqrt{
    \frac{1}{N} \sum_{i=1}^N
    \left[V_k^{\rm (A)}(\nu_i)-V_k^{\rm (S)}(\nu_i)\right]^2
  },
\label{eq:618}
\end{eqnarray}
where superscript (A) stands for ``analytic'' and (S) stands for
``simulation'', and $V_k({\rm max})$ is the maximum value of a MF
$V_k$ evaluated by the Gaussian formula, i.e., $V_k({\rm max}) =
V_0(\nu=-\infty) = 1$, $V_1(\nu=0)$, $V_2(\nu=1)$ and $V_3(\nu=0)$ for
$k = 0$, $1$, $2$, $3$, respectively. The set of various thresholds
$\nu_i$ are given by choosing $N=121$ equally spaced points with
interval $\Delta\nu=0.05$ in the range $-2\leq\nu\leq+4$.

When the mean differences defined above are within errors of the
numerical simulation, the indicated differences are not distinguished
from real differences between analytic predictions and true values of
MFs. Therefore, we also define the mean errors of the normalized MFs
in simulation data,
\begin{eqnarray}
  E_k \equiv
  \frac{1}{V_{k}({\rm max})}
  \sqrt{\frac{1}{N}
  \sum_{i=1}^N
  \left[\Delta V_k^{({\rm S})}(\nu_i)\right]^2},
\label{eq:619}
\end{eqnarray}
and the chi-square per degree of freedom:
\begin{eqnarray}
  \widehat{\chi}^2 =
  \frac{1}{N}
  \sum_{i=1}^N
  \frac{
    \left[V_k^{(\rm{A})}(\nu_i) - V_k^{(\rm{S})}(\nu_i)\right]^2}
      {\left[\Delta V_k^{(\rm{S})}(\nu_i)\right]^2}, 
\label{eq:620}
\end{eqnarray}
where $\Delta V_k^{(\rm{S})}(\nu_i)$ indicate the 1$\sigma$ errors in
the simulation data. When $D_k \simlt E_k$, the indicated differences
are within the errors originated from a numerical resolution of the
simulation, and are not considered as real differences. In fact, this
criterion is consistent with a usual chi-square criterion
$\widehat{\chi}^2 \simlt 1$. When $\widehat{\chi}^2 \simgt 1$, the
differences are interpreted as deviations of the analytic predictions
from true values of MFs. Thus accuracies of the analytic predictions
are indicated by the quantity $D_k$ when $\widehat{\chi}^2 \simgt 1$.
On the other hand, when $\widehat{\chi}^2 \simlt 1$, the statistic
$D_k$ represent the upper limits of the accuracies.

In Table~\ref{tab:64}, the above statistics are shown for all MFs. 
\begin{table}
\begin{center}
\caption{Values of $D_k$ of equation (\ref{eq:618}), $E_k$ of
equation (\ref{eq:619}), and $\widehat{\chi}^2$ of equation
(\ref{eq:620}).
\label{tab:64}}
\begin{tabular}{cccccccc}
%\phantom{.}
%\startdata
\hline\hline
& & \multicolumn{6}{c}{Smoothing Length $R (h^{-1}{\rm Mpc})$} \\
$k$ & {} & 10 & 15 & 20 & 25 & 40 & 50 \\ \hline
{}& $D_k^{\rm (2nd)}$ & 0.27\% & 0.050\% & 0.016\% & 0.010\% & 0.013\%
& 0.017\% \\ 
{}& $D_k^{\rm (linear)}$ & 1.58\% & 0.70\% & 0.38\% & 0.24\% & 0.14\%
& 0.15\% \\ 
$0$ & $E_k$ & 0.003\% & 0.003\% & 0.005\% & 0.010\% & 0.030\% &
0.056\% \\ 
{}& $\widehat{\chi}^2_{\rm 2nd}$ & 3631\phantom{\phd000} &
34.76 & 5.950 & 1.371 & 0.314 & 0.366 \\ 
{}& $\widehat{\chi}^2_{\rm lin}$ & 13317\phantom{\phd000} &
278.9 & 66.29 & 19.81 & 3.608 &
2.850 \\ \hline 
{}& $D_k^{\rm (2nd)}$ & 7.6\% & 3.0\% & 1.8\% & 1.6\% & 2.0\% & 1.8\% \\
{}& $D_k^{\rm (linear)}$ & 12\% & 6.9\% & 5.2\% & 5.2\% & 5.1\% &
4.4\% \\ 
$1$ & $E_k$ & 0.75\% & 0.73\% & 0.99\% & 1.9\% & 5.5\% & 6.3\% \\
{}& $\widehat{\chi}^2_{\rm 2nd}$ & 723.9 &
31.22 & 5.284 & 1.088 & 0.136 & 0.163 \\ 
{}& $\widehat{\chi}^2_{\rm lin}$ & 6678 & 180.2 & 60.63 & 15.60 &
1.610 & 1.072 \\ \hline 
{}& $D_k^{\rm (2nd)}$ & 14\% & 6.7\% & 4.0\% & 3.1\% & 3.8\% & 6.1\% \\ 
{}& $D_k^{\rm (linear)}$ & 25\% & 18\% & 13\% & 11\% & 9.1\% & 9.6\% \\ 
$2$ & $E_k$ & 1.6\% & 2.1\% & 3.0\% & 4.3\% & 9.2\% & 15\% \\ 
{}& $\widehat{\chi}^2_{\rm 2nd}$ & 148.1 & 24.71 & 3.547 & 0.775 &
0.204 & 0.225 \\ 
{}& $\widehat{\chi}^2_{\rm lin}$ & 3146 & 128.2 & 49.34 & 12.10 &
1.848 & 1.066 \\ \hline  
{}& $D_k^{\rm (2nd)}$ & 14\% & 7.1\% & 4.5\% & 3.9\% & 5.4\% & 7.7\% \\ 
{}& $D_k^{\rm (linear)}$ & 18\% & 13\% & 10\% & 8.9\% & 9.3\% & 11\% \\ 
$3$ & $E_k$ & 2.3\% & 3.4\% & 4.8\% & 6.1\% & 1.0\% & 14\% \\
{}& $\widehat{\chi}^2_{\rm 2nd}$ & 265.2 & 14.70 & 2.164 & 0.601 &
0.384 & 0.297 \\  
{}& $\widehat{\chi}^2_{\rm lin}$ & 841.0 & 41.67 & 13.06 & 5.043 &
1.513 & 0.784 \\ \hline  
%\enddata
\end{tabular}
\end{center}
\end{table}
In the Table, $D_k^{\rm (2nd)}$ and $D_k^{\rm (lin)}$ indicate the
equation (\ref{eq:618}) for 2nd-order and linear perturbation theory,
respectively. Likewise, $\widehat{\chi}^2_{\rm 2nd}$ and
$\widehat{\chi}^2_{\rm lin}$ indicate the equation (\ref{eq:620}) for
2nd-order and linear perturbation theory, respectively.

Overall, since $D_k^{\rm (2nd)}$ are smaller than $D_k^{\rm (lin)}$ in
each case, the second-order predictions show better agreement with
simulation data. For example, a prediction of the second-order theory
of $V_3$ agrees with the simulation data by $4\%$ accuracy level,
while that of the linear theory agrees by $10\%$ level. In other
words, the chi-square statistics of the second-order theory are always
smaller than that of the linear theory. Since the linear theory is
less accurate approximation than the second-order theory, this is a
natural consequence. When the smoothing lengths are large, the
differences between second-order predictions and the simulation data
are not distinguishable. There are two reasons for this tendency.
First, the analytic predictions are more accurate on large scales.
Second, the simulation data suffers more errors by finite volume
effects.

On smaller scales, $R\sim 10\himpc$, the analytic predictions of the
perturbation theories are less accurate because of the nonlinear
evolutions. Even in this regime, the second-order theory is more
accurate than the linear theory. Therefore, the predictions of the
second-order theory is always better than that of the linear theory in
the regime we investigate, i.e., $10\himpc \simlt R \simlt 50\himpc$.

At this point, one should note that there are correlations between the
adjacent thresholds $\nu_i$ and $\nu_{i+1}$. If the value of the MFs
increases (decreases) at some threshold density $\nu_i$, there is
higher possibility that the value also increases (decreases) at the
next threshold density $\nu_{i+1}$ because of the correlation. This is
particularly prominent in the Euler characteristic $V_3$. Since the
chi-square analysis cannot take such correlations into account, the
discrepancies between the theory and the data might be over
emphasized.

\subsection{The comparison of the MFs in terms of the rescaled
thresholds}

\begin{figure}
\epsscale{0.9}
\begin{center}
\plotone{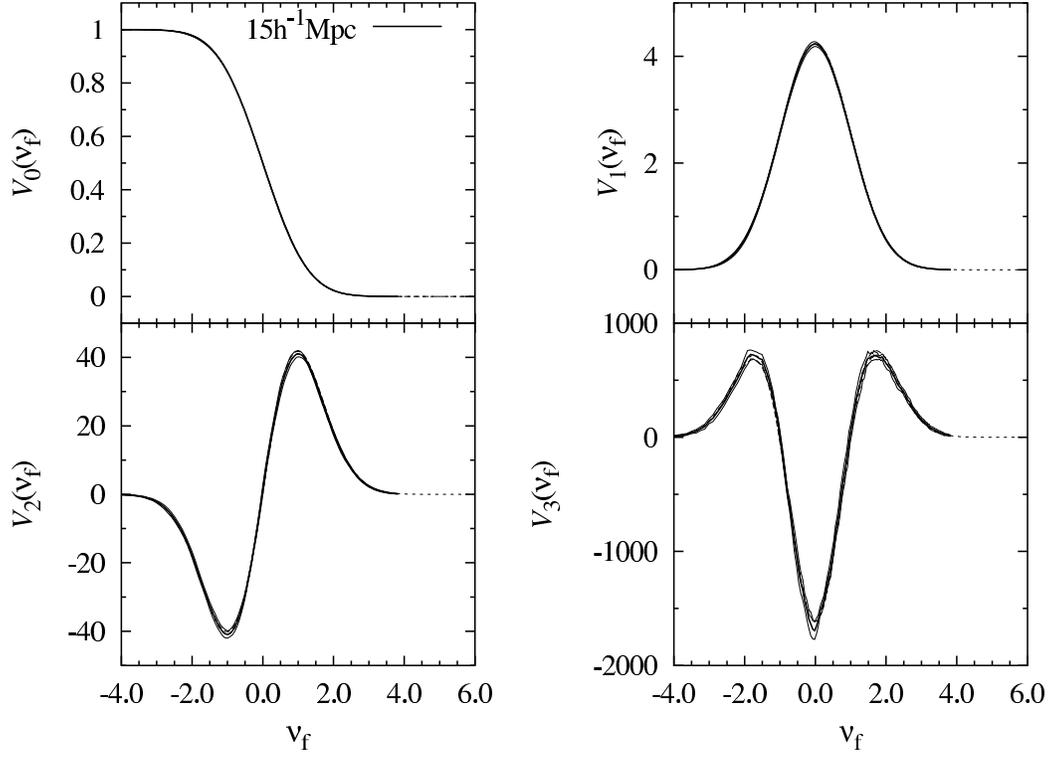}
\caption{Same as Figure \ref{fig:6100}, but for the threshold density 
$\nu_{\rm f}$ rescaled by volume fraction.
}
\label{fig:6101}
\end{center}
\end{figure}

\begin{figure}
\epsscale{0.9}
\begin{center}
\plotone{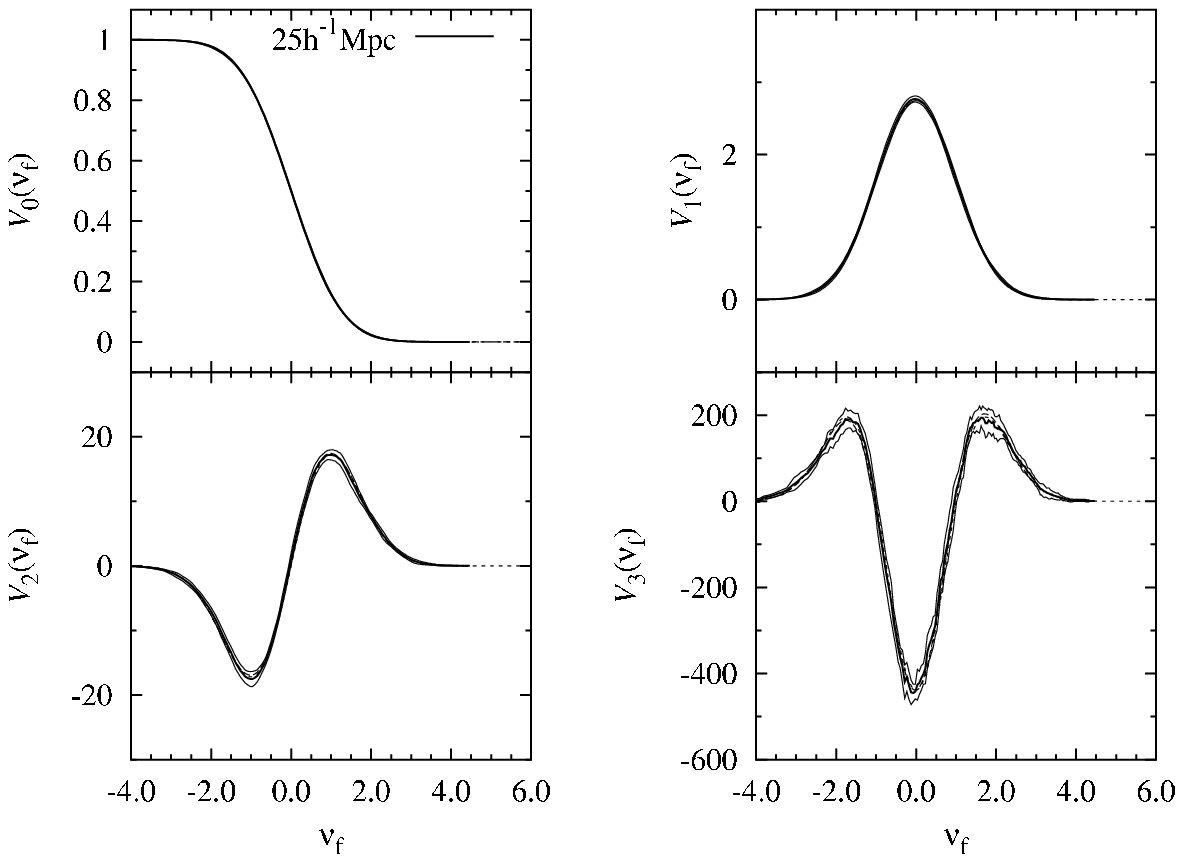}
\caption{Same as Figure \ref{fig:6100}, but for the smoothing length 
$R=25h^{-1}{\rm Mpc}$ and the threshold density $\nu_{\rm f}$ rescaled by volume 
fraction.
}
\label{fig:6104}
\end{center}
\end{figure}

\begin{figure}
\epsscale{0.9}
\begin{center}
\plotone{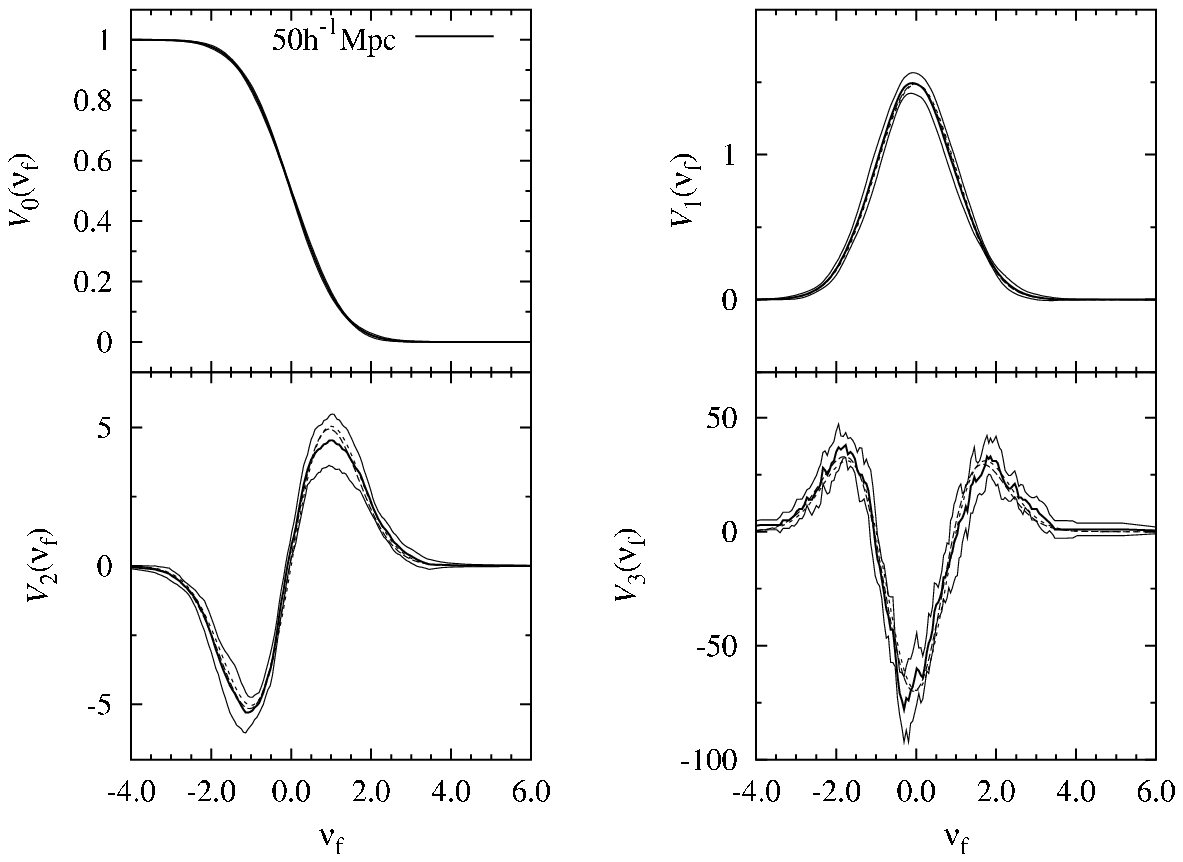}
\caption{Same as Figure \ref{fig:6100}, but for the smoothing length 
$R=50h^{-1}{\rm Mpc}$ the threshold density $\nu_{\rm f}$ rescaled by volume 
fraction.
}
\label{fig:6105}
\end{center}
\end{figure}

\begin{figure}
\begin{center}
\epsscale{0.9}
\plottwo{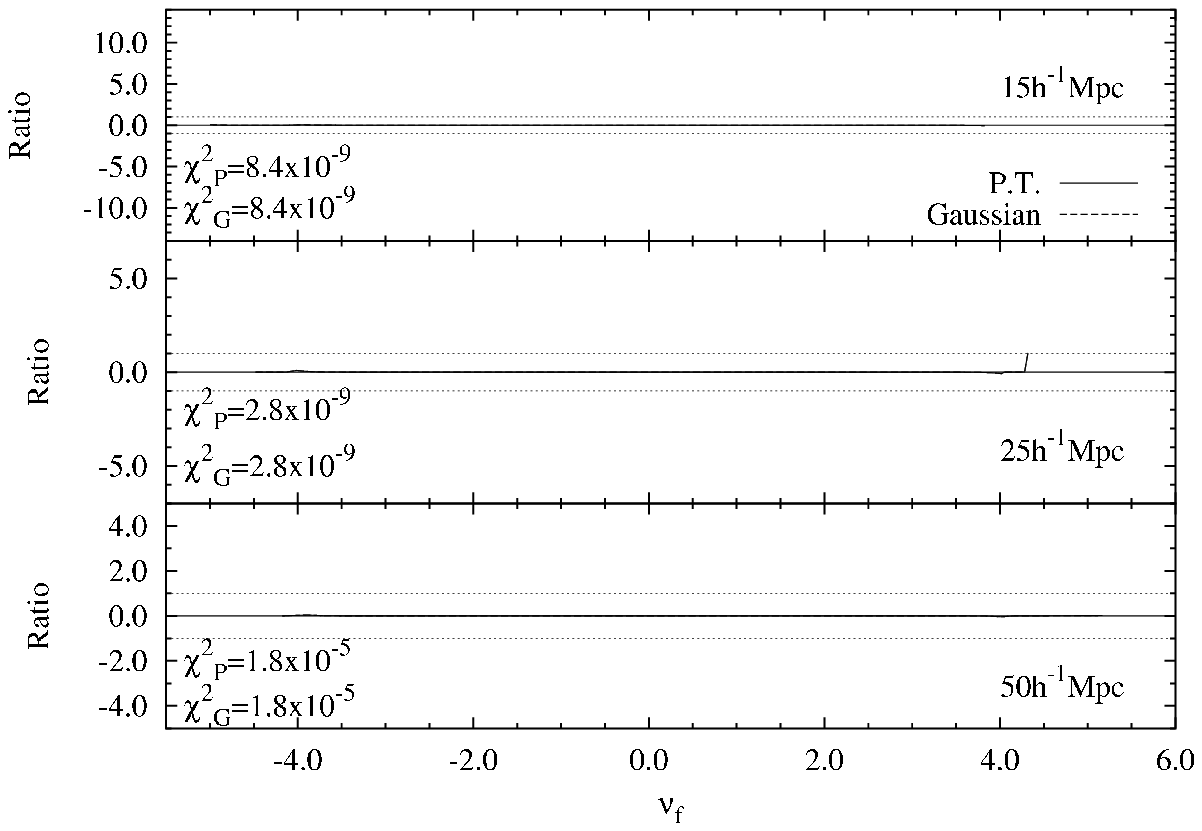}{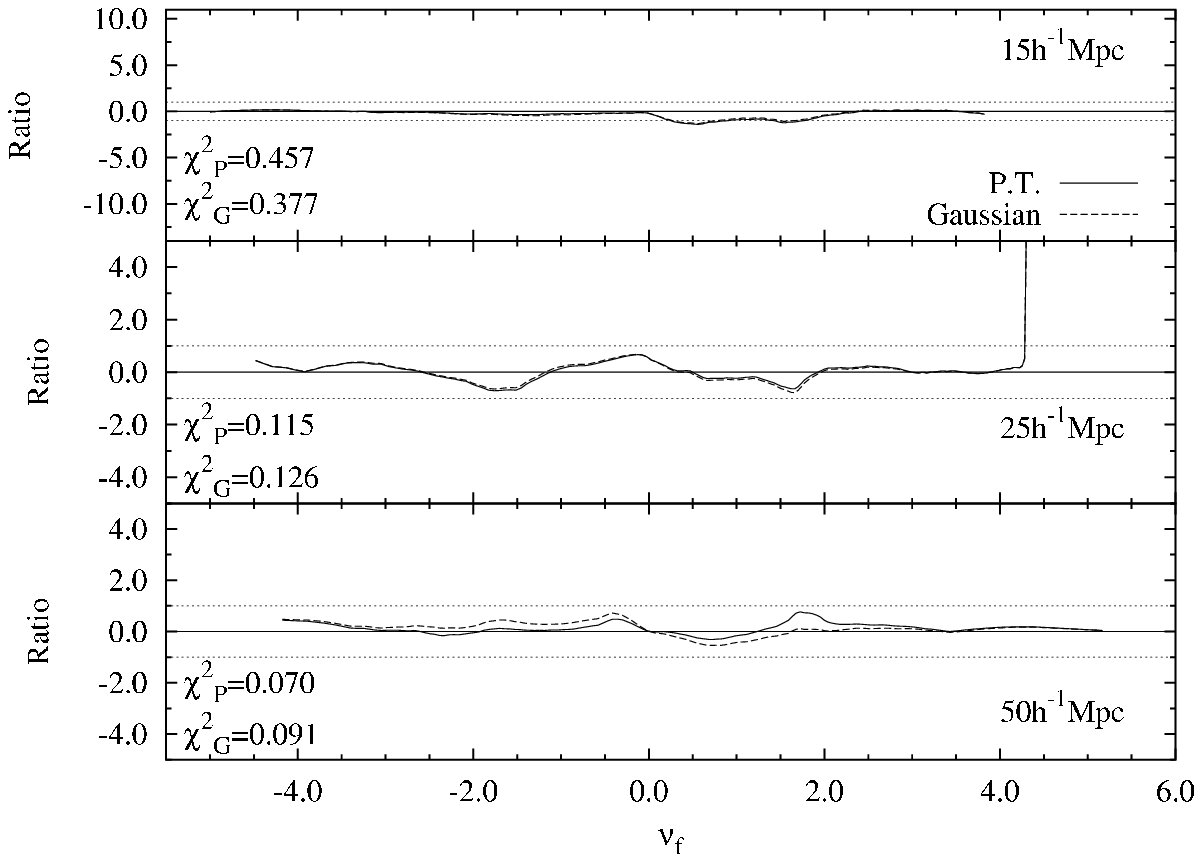}
\plottwo{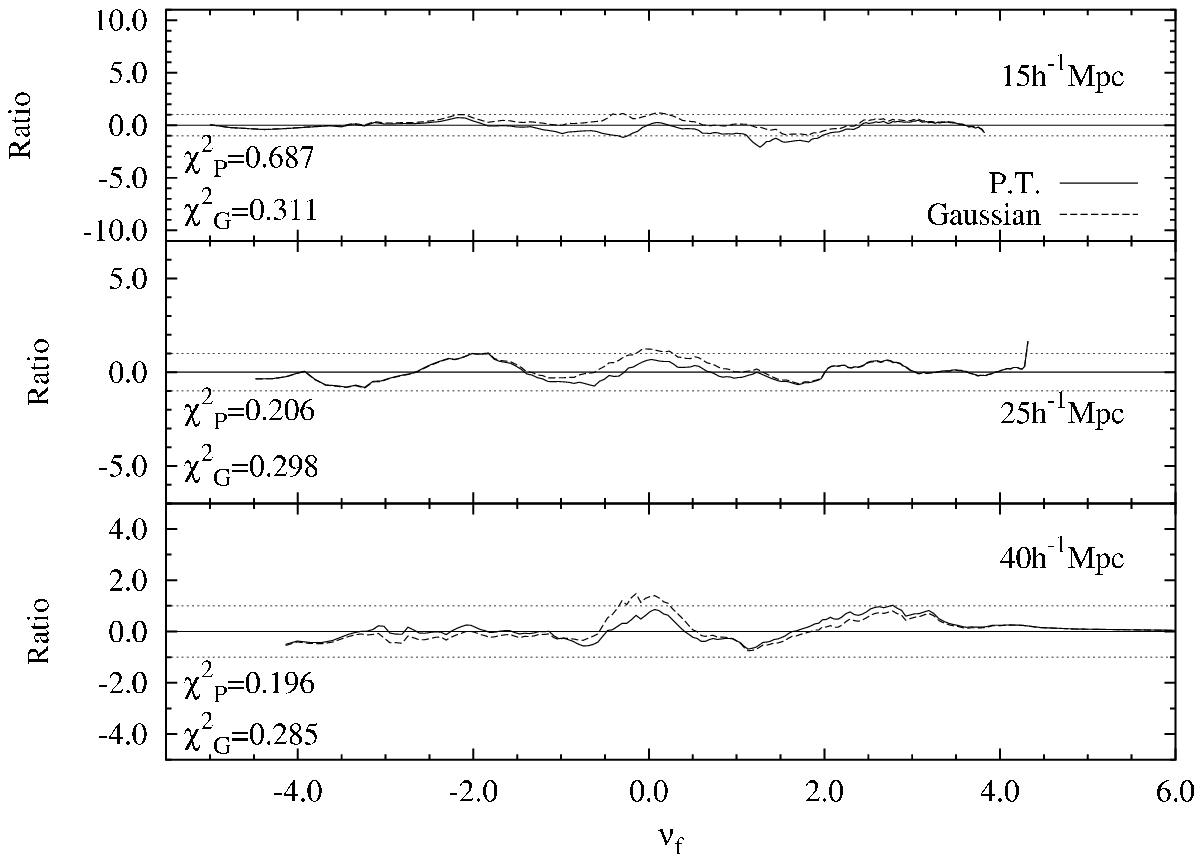}{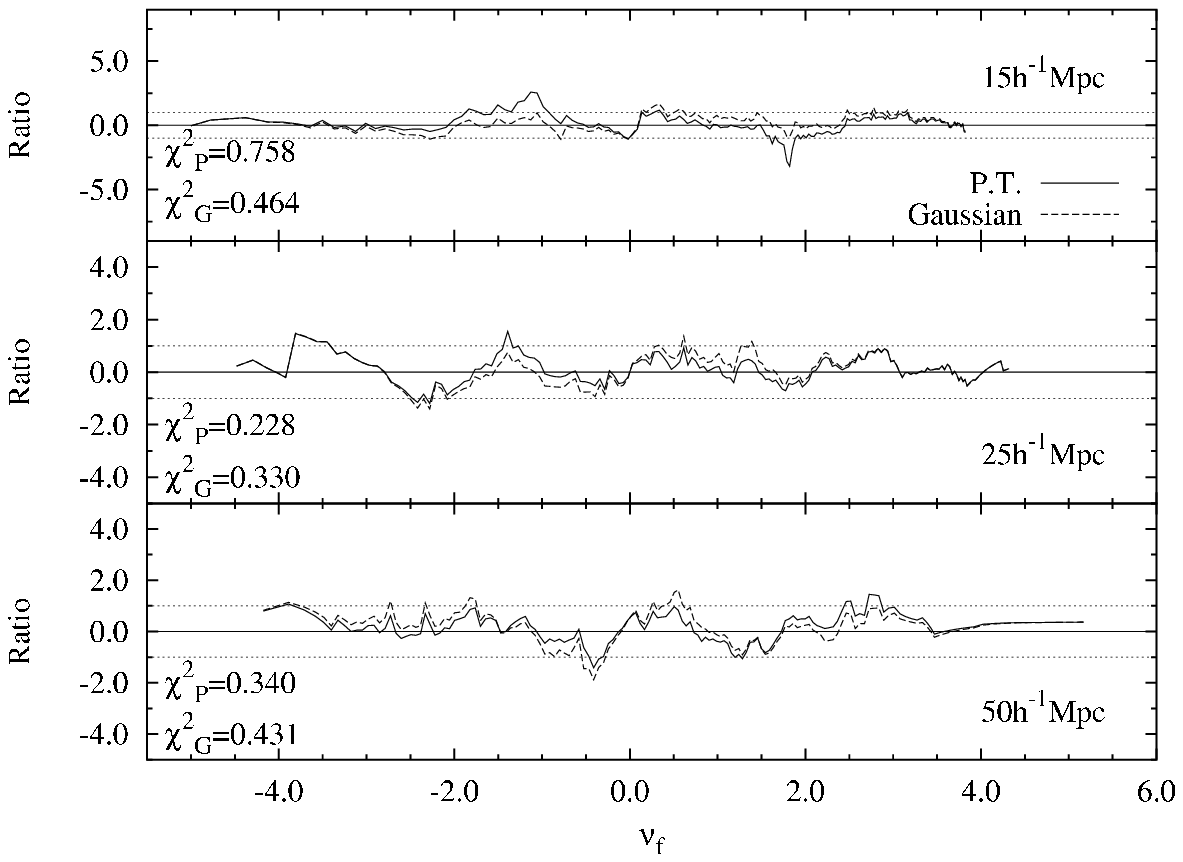}
\caption{Same as Figure \ref{fig:61}, but for the threshold density 
$\nu_{\rm f}$ rescaled by volume fraction.}
\label{fig:610}
\end{center}
\end{figure}

In the above comparisons, we use the threshold $\nu =
\delta/\sigma_0$. In the following, we use the rescaled threshold
$\nu_{\rm f}$ and make the same comparisons. Figures \ref{fig:6101},
\ref{fig:6104} and \ref{fig:6105} correspond to Figures
\ref{fig:6100}, \ref{fig:6102} and \ref{fig:6103}, respectively.
The
differences of the MFs between the analytic formulae and the
simulation are also shown in Figure \ref{fig:610}.

From these Figures, we find that the differences in terms of $\nu_{\rm
f}$ are smaller than those in terms of $\nu$ for all MFs.
Mathematically, one of the reason for this is that the skewness
parameters only appear as combinations of the form $S^{(a)}-S^{(0)}$
in the analytic formulae (eq. [\ref{eq:435}]), and the three values of
skewness parameters in CDM cosmological model are close. Since
$S^{(1)}-S^{(0)}$ is almost zero in broad range of the smoothing
length, the deviations from analytic formulae are mainly depend on the
value of $S^{(2)}-S^{(0)}$.

The differences $D_k$, errors $E_k$, and chi-square per d.o.f.
$\widehat{\chi}^2$ are also calculated in terms of $\nu_{\rm f}$.
These statistics are shown in Table \ref{tab:65}. 
\begin{table}
\begin{center}
\caption{Same as Table
\ref{tab:64}, but for the rescaled threshold $\nu_{\rm f}$.
\label{tab:65}}
\begin{tabular}{cccccccc}
\hline\hline
& & \multicolumn{6}{c}{Smoothing length $R (h^{-1}{\rm Mpc})$} \\
$k$ & {} &10 & 15 & 20 & 
25 & 40 & 50 \\ \hline
{} & $D_k^{\rm (2nd)}$ & 1.4\% & 0.45\% & 0.35\% & 0.64\% & 1.5\% & 1.3\% \\
{} & $D_k^{\rm (linear)}$ & 1.2\% & 0.42\% & 0.39\% & 0.69\% & 2.1\% &
2.4\% \\ 
$1$ & $E_k$ & 0.75\% & 0.73\% & 0.99\% & 1.9\% & 5.5\% & 6.3\% \\
{} & $\widehat{\chi}^2_{\rm 2nd}$ & 4.441 & 0.457 & 0.121 & 0.115 &
0.083 & $0.070$ \\ 
{} & $\widehat{\chi}^2_{\rm lin}$ & 1.917 & 0.377 & 0.140 & 0.126 &
0.113 & $0.091$ \\ \hline 
{} & $D_k^{\rm (2nd)}$ & 3.4\% & 1.4\% & 1.4\% & 1.9\% & 3.4\% & 5.5\%
\\
{} & $D_k^{\rm (linear)}$ & 1.5\% & 1.2\% & 1.8\% & 2.5\% & 4.3\% &
6.1\% \\ 
$2$ & $E_k$ & 1.6\% & 2.1\% & 3.0\% & 4.3\% & 9.2\% & 1.5\% \\
{}& $\widehat{\chi}^2_{\rm 2nd}$ & 7.397 & 0.687 & 0.248 & 0.206 &
0.221 & 0.196 \\ 
{}& $\widehat{\chi}^2_{\rm lin}$ & 0.961 & 0.311 & 0.303 & 0.298 &
0.293 & 0.285 \\ \hline 
{}& $D_k^{\rm (2nd)}$ & 3.2\% & 2.1\% & 2.3\% & 2.6\% & 5.3\% & 7.5\% \\
 {}& $D_k^{\rm (linear)}$ & 1.4\% & 2.2\% & 3.1\% & 3.5\% & 6.5\% & 88\% \\
$3$ & $E_k$ & 2.4\% & 3.6\% & 5.0\% & 6.2\% & 11\% & 15\% \\
{}& $\widehat{\chi}^2_{\rm 2nd}$ & 6.115 & 0.758 & 0.280 & 0.228 &
0.519 & 0.340 \\ 
{}& $\widehat{\chi}^2_{\rm lin}$ & 0.351 & 0.464 & 0.408 & 0.330 &
0.599 & 0.431 \\ \hline 
%\enddata
\end{tabular}
\end{center}
\end{table}
It is seen that the
differences $D_k$ are within several percent in almost all MFs. Most
of the $\widehat{\chi}^2$ are less than 1 and the prediction of the
analytic formulae are not distinguishable with the simulation results.
Interestingly, the second-order predictions of $V_1$, $V_2$ and $V_3$
for $R = 10\himpc$ are less accurate than the linear predictions. On
this scale, both the linear theory and the second-order theory can be
inaccurate. A possible interpretation is that the strongly nonlinear
dynamics affect the MFs to cancel the second-order correction term in
the analytic formulae.

As a result, the linear theory and the second-order theory are
difficult to be distinguished when rescaled threshold $\nu_{\rm f}$ is
used. This is a good news for testing the primordial non-Gaussianity in
the density fluctuations since the nonlinear dynamics in course of the
density evolution does not significantly alter the shape of MFs in
Gaussian fields.

\section{Conclusions and Discussion}

We analyze the Minkowski functionals using a large $N$-body simulation
of a $\Lambda$CDM model with a box size of $(1024h^{-1}{\rm Mpc})^3$
and canonical cosmological parameters, $\Omega_0=0.3$,
$\lambda_0=0.7$, $h=0.6667$, $\Gamma=0.2$. The detailed analysis of
the MFs with such a large $N$-body simulation is unprecedented. The
validity levels and regions of the analytic formulae of the MFs are
studied in detail. We focus on the transition scales from linear to
nonlinear evolution, 10--$50h^{-1}{\rm Mpc}$, and calculate the
typical differences of the MFs between the analytic formulae and the
$N$-body simulation.

The variance parameters $\sigma_0$, $\sigma_1$, and the skewness
parameters $S^{(a)}(a=0,1,2)$ are calculated from the simulation and
compared with analytic predictions. It is found that
$S^{(a)}(a=0,1,2)$ agree with the analytic values within the
simulation errors in a large range of smoothing lengths. In the
previous work of such comparisons, physical sizes of the simulations
are not large enough as in this work. Thus the skewness parameters
from the simulation suffer from the cosmic variance and disagree with
analytic predictions. It is a new result that the skewness parameters
calculated from such a large $N$-body simulation as $(1024h^{-1}{\rm
Mpc})^3$ agree with the analytic predictions. On the other hand, the
variance $\sigma_0$ calculated from the $N$-body simulation is
slightly smaller than the linear prediction naturally expected in
weakly non-linear regime.

In this paper we use two definitions of the threshold density, i.e.,
$\nu$ defined by the value of the density field, and $\nu_{\rm f}$
defined by the volume fraction. Nonlinear effects on MFs are stronger
against the threshold $\nu$ than $\nu_{\rm f}$. In most of the
previous studies on the topology of the large-scale structure, the
threshold density $\nu_{\rm f}$ has been adopted. The reason for this
choice is that shape of the genus curve is empirically known not to be
much affected by the nonlinear evolution. This was also analytically
shown in weakly nonlinear regime. In this paper, it is shown that the
deviations of the MFs on nonlinear scales are indistinguishable with
Gaussian predictions even though we use a large $N$-body sample. To
detect the effects of the weakly nonlinear evolution in $\nu_{\rm f}$
case, a larger sample with less cosmic variance will be needed.
Sampling the galaxies in over $1(\higpc)^3$ comoving volume is within
reach of the future surveys. For example, in the SDSS Luminous Red
Galaxy survey \citep{eisenstein01}, the large-scale structure up to $z
\simlt 0.5$ is being probed. The covered volume is approximately
$1(\higpc)^3$ in this survey. Therefore, our estimates of the errors
of MFs in the simulation roughly correspond to the errors in this kind
of surveys. If we have still larger surveys, e.g., a $z\simlt 1$
survey with a large sky coverage, the MFs can be measured in exquisite
detail. When some non-Gaussian behavior in MFs is ovservationally
found, it is important to distinguish the non-Gaussianity of the
primordial density field from that caused by non-linear evolution.
This work gives useful information in this respect.

In this work, we use the density distribution in real space. In
redshift surveys, the observed distributions of galaxies are in
redshift space. The shape of MFs as functions of the threshold is
unaffected by peculiar velocity effect in redshift space in linear
theory \citep{matsubara96}. Analytic formula of MFs in redshift space
with nonlinear effect has not been derived yet. In future work, it
will be interesting to investigate weakly nonlinear effect on MFs in
redshift space, both analytically and numerically.

\acknowledgements

TM acknowledges support from the MEXT, Grant-in-Aid for Encouragement
of Young Scientists, 15740151, 2003. YPJ is supported in part by
NKBRSF(G19990754) and NSFC.

\end{document}